\documentclass[aps,pre,reprint,groupedaddress]{revtex4-1}
\usepackage[english]{babel}
\usepackage{amsmath}
\usepackage{graphicx}
\usepackage{dcolumn}
\usepackage{bm}
\usepackage{epstopdf}

\usepackage{epsfig}

\newcommand{\nbini}{{n_i}} 
\newcommand{\nbinj}{{n_j}} 

\renewcommand{\th}{{}} 
\newcommand{\bin}{{\rm bin}}

\newcommand{\mNbar}{{\overline m}_N}

\newcommand{\scat}{{\rm sc}}

\newcommand{\bec}{{\rm bec}}

\newcommand{\myskip}[1]{}

\renewcommand{\d}{{\rm d}}

\newcommand{\BEQ}{\begin{eqnarray}}
\newcommand{\EEQ}{\end{eqnarray}}
\newcommand{\BEA}{\begin{eqnarray}}
\newcommand{\EEA}{\end{eqnarray}}
\newcommand{\nn}{\nonumber}
\newcommand{\Sigmab}{\overline\Sigma}

\newcommand{\cm}{{\rm cm}}
\newcommand{\gr}{{\rm gr}}
\newcommand{\km}{{\rm km}}

\newcommand{\s}{{\rm s}}
\newcommand{\p}{\partial}

\newcommand{\kpc}{{\rm kpc}}

\newcommand{\eV}{{\rm eV}}
\newcommand{\MeV}{{\rm MeV}}

\newcommand{\Gal}{{\it G}}

\newcommand{\LCDM}{{$\Lambda$CDM}}

\newcommand{\MACHO}{{M\small{ACHO}}}
\newcommand{\Macho}{{M\small{ACHO}}}
\newcommand{\WIMP}{{W\small{IMP}}}

\newcommand{\ALP}{{A\small{LP}}}
\newcommand{\DM}{{D\small{M}}}
\newcommand{\BH}{{B\small{H}}}
\newcommand{\PBH}{{P\small{BH}}}
\newcommand{\BEC}{{B\small{EC}}}

\begin{document}

\title{Subjecting dark matter candidates to the cluster test }


\author{Theodorus Maria Nieuwenhuizen$^{1,2}$}

\address{ 
$^1$Institute for Theoretical Physics,  University of Amsterdam,  Science Park 904, 1090 GL  Amsterdam, The Netherlands \\
$^2$International Institute of Physics, Federal University of Rio Grande do Norte, Natal, Brazil}

\begin{abstract}
Galaxy clusters, employed by Zwicky to demonstrate the existence of dark matter, pose new stringent tests. 
If merging clusters demonstrate that dark matter is self-interacting with cross section $\sigma/m\sim 2\,\cm^2/\gr$, 
MACHOs, primordial black holes and light axions that build \MACHO s are ruled out as cluster dark matter.
Recent strong lensing and X-ray gas data of the quite relaxed and quite spherical cluster A1835 allow to test the cases of dark matter 
with  Maxwell-Boltzmann, Bose-Einstein and Fermi-Dirac  distribution, next to Navarro-Frenck-White profiles.
Fits to all these profiles are formally rejected at over $5\sigma$, except in the fermionic situation.
The interpretation in terms of (nearly) Dirac neutrinos with mass of $1.61^{+0.19}_{-0.30}$ eV/$c^2$
is consistent with results on the cluster A1689,  with the WMAP, Planck and DES dark matter fractions 
and with the nondetection of neutrinoless double $\beta$-decay. The case will be tested in the 2018 KATRIN experiment.
\end{abstract}

\maketitle

\section{Introduction}

The existence of dark matter (\DM), or some equivalent effect, is beyond doubt and proves the existence of new degrees of freedom. 
The usual suspects are WIMPs, axions and sterile neutrinos.
The standard model of cosmology \LCDM \, explains Big Bang Nucleosynthesis (BBN), the Cosmic Microwave Background (CMB) 
and the Baryon Acoustic Oscillations (BAO), as recently supported by the Dark Energy Survey  \cite{troxel2017dark}. 
But there are several issues, such as:
The DM particle has been sought intensly but not found  \cite{arcadi2017waning}, neither is there a hint for supersymmetry at the LHC.
BBN faces the $^7$Li problem  \cite{cyburt2016big}, the CMB has a small Hubble constant  \cite{freedman2017cosmology}
and faces foreground issues \cite{verschuur2016nature,vavryvcuk2017missing}.
 Red-and-dead galaxies require early structure formation \cite{croton2008red}, as does a dusty  galaxy at $z\sim 7$ with some 
$3\times 10^{11}M_\odot$ in gas \cite{strandet2017ism}.
 Lyman-$\alpha$ clouds are supposed to be stabilized by a high temperature plasma, which should be easy to detect but never was.

These and other sobering results motivate to reconsider other \DM \ options,
like primordial black holes (\PBH s) or  \MACHO  \ dark matter. 
\PBH s were thought to be ruled out,  but became fashionable again after the discovery of gravitational waves from \BH \ mergers,
to meet fresh criticism \cite{koushiappas2017dynamics}. 
A \MACHO \ can be e.g. a planet or a solar mass object,
that may consist of normal matter, but also stand for a self-gravitating Bose-Einstein condensate (\BEC) of axions or axion-like particles (\ALP s).
From another angle, our studies of lensing by the cluster A1689 consistently yield good
fits for neutrino DM  \cite{nieuwenhuizen2009non,nieuwenhuizen2013observations,nieuwenhuizen2016dirac}.

Supposing that \DM \ does not exist but that Newton's law gets modified below a critical value of the acceleration
has been fruitful for the description for galactic rotation curves \cite{milgrom1983modification}.
However, it has been demonstrated that these theories, in particular M\small{OND}, Emergent Gravity, $f(R)$ and M\small{OG},
run into serious troubles for galaxy clusters.
The fairly relaxed cluster Abell 1689  posed problems for these theories  \cite{nieuwenhuizen2017zwicky},
as did a second relaxed cluster,  A1835 \cite{nieuwenhuizen2017modified}.
To function in clusters, MOND and EG would need additional \DM, e.g.,  in the form of $\sim 2$ eV thermal neutrinos.
This hot \DM \  is known to induce free streaming in the early Universe, thus suppressing structure formation.
They are considered as ruled; in fact the sum of neutrino masses
is estimated to lie in the 0.1 -- 0.3 eV range. 
Nevertheless, a rarely considered question is: has structure formation indeed been linear?

With the road for non-Newtonian gravity essentially closed in our contribution to FQMT'15 \cite{nieuwenhuizen2017zwicky}, 
the way forward is to study implications  of particle dark matter 
theories in galaxy clusters. In contrast to CMB and BAO theories, relaxed clusters have simple physics:
one may assume that some kind of equilibrium has been reached, so that the history needs not be considered.
As such, they put important bench marks.


The paper is composed as follows.
In section 2 we consider the effect of DM self-interaction.
In section 3 we discuss data for the cluster A1835 and their binning.
This is applied to NFW fits in section 4 and to thermal fits in section 5.
The paper ends with a summary and an outlook. Throughout the paper we use the reduced Hubble constant $h=0.7$.

\section{On dark matter self-interaction }

\subsection{\MACHO s and \PBH s}

In clusters there are too few baryons to account for all the  \DM \, but \MACHO s may consist of axions or \ALP s, or be \PBH s.
Let us look at a specific cluster, the ``train wreck'' cluster Abell 520. 
It reflects  the past collision of at least three sub-clusters, which are on their exit.
Surprisingly, it has a central starless core of a few times $10^{13}M_\odot$ and mass-to-light ratio 
860 $M_\odot/L_\odot$  \cite{jee2012study,clowe2012dark,jee2014hubble}.
This has been modelled by self-interacting DM (SiDM) with an elastic scattering cross section of $\sigma/m\sim 1.7 \, \cm^2/\gr$.  
A similar estimate comes from the Bullet Cluster  \cite{markevitch2004direct}.
\Macho s and  \PBH s can not have this;  for 1 Earth mass, e. g.,   they would need the gigantic value $\sigma\sim 50\,{\rm AU}^2$. 
If SiDM exists, \MACHO s  are ruled out as the cluster \DM.
For both clusters the existence of SiDM  has been questioned, however \cite{robertson2016does,peel2017sparse}.
But also the cluster A3827 yields a mild indication for self-interaction, 
($\sigma/m)$$\times$ $\cos i = 0.68^{+0.28}_{-0.29} \cm^2/\gr$,
where $i$ is an inclination angle   \cite{massey2017dark}.

\subsection{\WIMP s}

The same argument applies  to \WIMP s, though in a much weaker form.
Intuitively, scattering occurs by contact interaction if particles come within their Compton radius. The condition
$\sigma_\scat< ({\hbar}/{mc})^2$ then leads to
\BEQ
m\lesssim \Big(\frac{\hbar^2\cm^2}{2c^2\gr}\Big)^{1/3}= 40 \, \frac{\MeV}{c^2},
\EEQ
which would explain why no WIMP has been observed in the GeV regime.
To go beyond this puts a constraint on theories.

\subsection{Sterile neutrinos}
In recent years attention has been payed to sterile neutrinos, so-called warm \DM.
In particular the report of a 3.5 keV $\gamma$-ray line, possibly related to a 7 keV sterile neutrino, has been
 inspiring  \cite{bulbul2014detection,boyarsky2014unidentified}. 
For elastic scattering the value $\sigma/m\sim 2\,\cm^2/\gr$ may not look problematic, 
but actually they should hardly interact at all, since
 sterile-sterile neutrino scattering happens indirectly via their mixing with standard `active' neutrinos.
For an active-sterile mixing angle $\theta_{14}$, 
the cross section can be estimated as $\sigma\sim \theta_{14}^8G_F^2(\hbar m_ec^3)^2= \theta_{14}^8\,1.4\,10^{-44}\,\cm^2$
\cite{lesgourgues2013neutrino}.
With $m=7.02$ keV and $\sin^22\theta_{14}= 0.69-2.29 \,10^{-10}$  \cite{cappelluti2017searching}
 it follows that $\sigma_{}/m=10^{-37}-10^{-36}\,\cm^2/\gr$.
If sterile neutrinos are to make up SiDM, they need an another, strong scattering mechanism.

\subsection{Axions and axion-like particles} 

 ALPs may be as light as $10^{-22}$ eV; with eV masses they will be thermal; 
if heavier,  they act as WIMPs.  Light ones may form Bose-Einstein condensates (\BEC s).
It has been proposed that very light ones, $m\sim 10^{-22}$ eV, build \BEC s which act as MACHOs  \cite{hui2016hypothesis}.
However, \MACHO \  scenarios can not act as SiDM.

Let us see whether perhaps the whole cluster DM can be one Mpc-sized BEC constituted by \ALP s.
Its ground state wavefunction  satisfies the Schr\"odinger equation
\BEQ
-\frac{\hbar^2}{2m}\nabla^2\psi_0+m\varphi\psi_0(r)=E\psi_0(r),\quad
\EEQ
and the Poisson equation, which relates the gravitational potential $\varphi$ to  the mass density $\rho_G$ of the Galaxies, 
the $\rho_g$ of the X-ray gas and the $\rho_\DM$ of the DM, 
\BEQ \label{Poissoneqn}
\nabla^2\varphi=4\pi G\rho, \quad 
\rho=\rho_B+\rho_\DM ,\quad 
\rho_B=\rho_G+\rho_g.
\EEQ
Here $\rho_\DM=\psi_0^2$ with normalisation $\int\d^3r\psi_0^2=M_\bec$.
In the cluster centre the mass  density is known to stem mainly from the brightest cluster galaxy, so $\psi_0^2(0)\ll \rho_G(0)$.
Hence the potential is harmonic, $\varphi=\frac{1}{2}m\omega^2 r^2$. With $\rho_G(0)\sim 10^{12}$ $M_\odot/(10\,\kpc)^3$ it has a 
frequency $\omega\approx[4\pi G\rho_G(0)/3]^{1/2}\sim 1/10^7\,{\rm yr}$.
This problem is solved in every  quantum mechanics textbook.
Its characteristic  length 
\BEQ
\ell_0=\sqrt{\frac{\hbar}{m\omega}}=\frac{4\,10^{-12}}{\sqrt{mc^2/\eV}} \, \kpc
\EEQ
is tiny on the cluster scale, so the condensate must basically act as a point mass, maximally equal to
\BEQ
M_\bec<\rho_G(0)\ell^3\sim \frac{5}{(mc^2/\eV)^{3/2}}\,10^{-26} M_\odot,
\EEQ
which even for $m\sim 10^{-22}$ eV is less than $10^8M_\odot$ and thus negligible.
Extended DM distributions must thus have many BECs acting as \MACHO s, a scenario discussed already.
Hence light axions and \ALP s  are problematic as SiDM.

\section{A1835  data and their binning}

For the cluster A1835 theories of DM can be tested on recent data for $M_{2d}(r)$, the mass in a cylinder around the cluster centre  \cite{nieuwenhuizen2017modified}.
From the observed strong lensing arclets mass maps are generated; this being an underdetermined problem, 
an ensemble ${\cal N}=1001$ of compatible $2d$ mass maps is produced and from them their $M_{2d}$ values
at radii $r_n\sim a^n$ with $n=1,\cdots, 149$, such that  $(r_1,r_{149})=(4.03,1120)$ kpc.
In the centre only a few arclets occur, hence only $N=117$ of the $r_n$ contain data 
for $\Sigmab_n=\langle M_{2d}(r_n)\rangle/\pi r_n^2$ and their covariances 
$\Gamma_{mn}$   \cite{nieuwenhuizen2017modified}; the index $n=1,\cdots,N$ is relabelled accordingly.
The matrix $\Gamma$ has a big spread of eigenvalues, roughly between 
$0.5$ and $5\,10^{-15}\gr^2/\cm^4$.
The standard definition of $\chi^2$ involves $\Gamma^{-1}$ but small eigenvalues 
should  not matter and have to be regularised. Hereto we shall merely employ the data themselves.

As first step to eliminate the small eigenvalues, the $N$ data points are grouped in $N_\bin=17$ bins with in principle
$n_i=7$ points, but not all bins can be full.
 Choosing $n_8=5$ or $n_{10}=5$ we minimize bias around the bin 9, which has the smallest errors.
We can now  relabel the index   $n\to \{ ik \}$, according to the bin number $i=1,\cdots,N_\bin$ and the location $k=1,\cdots,n_i $ inside the bin;
this defines $r_{ik}$, $\Sigmab_{ik}$ and $\Gamma_{ik;jl}$.
As  bin centre $r_i$  we take   the geometrical average 
$ r_i=(\Pi_{k=1}^\nbini r_{ik})^{1/n_i}$.

As a new step, we divide out the theoretical value in the binning.
Given a theoretical or empirical $\Sigmab_\th(r)$, the data is binned as 
\BEQ \label{SiBbin} 
\Sigmab_i^\bin=\Sigmab_\th(r_i)\frac{1}{\nbini}\sum_{k=1}^{\nbini}  \frac{\Sigmab_{ik}}{\Sigmab_\th(r_{ik})}, 
\qquad i=1,\cdots, N_\bin. 
\EEQ
The standard binning with $\Sigmab_\th(r)\to 1$ would do less justice to the data than the best $\Sigmab_\th(r)$ fit,
and hence lead to  a loss of information. 
Moreover, the binning (\ref{SiBbin}) makes the choice of $r_i$ as good as any other. 
The binned covariances read
\BEQ \label{Gambin}
{\Gamma}_{ij}^\bin=\frac{\Sigmab_\th(r_i)\Sigmab_\th(r_j)}{\nbini \nbinj} \sum_{k=1}^{\nbini}\sum_{l=1}^{\nbinj }
\frac{\Gamma_{ik;jl}}{\Sigmab_\th(r_{ik})\, \Sigmab_\th(r_{jl})}. 
\EEQ
$\Gamma^\bin$ has eigenvalues typically from 0.07 to $5\,10^{-14}\gr^2/\cm^4$, hardly better than $\Gamma$.
The way to proceed is by noting that eq. (\ref{SiBbin}) puts forward a measure for the intra-bin fluctuations,
\BEQ 
\hspace{-6mm}
\gamma_i=\frac{\Sigmab^2_\th(r_i)}{\nbini^2}\sum_{k,l=1}^{\nbini} 
\Big | \Big ( \frac{\Sigmab_{ik}}{\Sigmab_\th(r_{ik})}-\frac{\Sigmab_i^\bin}{\Sigmab_\th(r_i)} \Big)  
 \Big ( \frac{\Sigmab_{il}}{\Sigmab_\th(r_{il})}-\frac{\Sigmab_i^\bin}{\Sigmab_\th(r_i)} \Big )\Big | .
 \EEQ
This is actually a square; without absolute values, it would vanish.
As final step, we add the $\gamma_i$  as diagonal regulator and define the total binned covariance matrix $C$, 
\BEQ \label{Cij=}
C_{ij}={\Gamma}_{ij}^\bin+\delta_{ij}\gamma_i.
\EEQ
The eigenvalues of $C$ go down to $\sim 10^{-7}$$\gr^2/\cm^4$, so 
further regularization with an ad hoc constant 
 $\delta\gamma_i=\gamma$ \cite{limousin2007combining,nieuwenhuizen2013observations,nieuwenhuizen2016dirac}
  is not needed. As measure for the goodness of the fit we take
\BEQ 
\label{X2Lim}
\chi^2(\Sigmab)= 
\sum_{i,j=1}^{N_\bin} \Big [\Sigmab_i^\bin-\Sigmab(r_i) \Big]\,
C^{-1}_{ij}  \Big[\Sigmab_j^\bin-\Sigmab(r_j)\Big] .
\EEQ
It differs from the standard $\chi^2$ in that the data and the covariances are binned employing the fit function $\Sigmab_\th(r)$.

To estimate the errors in fit parameters $p_1,p_2,\cdots$ we assume that the data involve Gaussian errors. 
Denoting $\Delta_i=\Sigmab_i^\bin-\Sigmab(r_i)$ and the errors by $\delta$,  the leading Gaussian errors  
of $\chi^2(\Sigmab)=\Delta C^{-1}\Delta$ are collected symbolically as
\BEQ \label{dX2Lim}
\delta\chi^2(\Sigmab)=
(\delta \Delta- \Delta C^{-1}\delta  C)C^{-1}(\delta \Delta-\delta  C\,C^{-1}\Delta),
\EEQ
where $\delta  \Delta_i=\sum_{k}(\p\Delta_i/\p p_k)\delta p_k$, and likewise for $\delta C_{ij}$. 
The covariances are defined from $\delta\chi^2(\Sigmab)\equiv \sum_{k,l}(X^{-1})_{kl}\delta p_k\delta p_l$ as
 $ \langle  \delta p_k\delta p_l \rangle  =X_{kl}$
and the errors in the $p_k$ as $\Delta p_k=(X_{kk})^{1/2}$.

\section{NFW fits}

 We first apply this to the Navarro-Frenk-White (NFW) profile  \cite{navarro1997universal},
\BEQ
\rho_{\rm NFW}=\frac{AR^3}{r (r+R)^{2}} =\frac{200c^3\rho_c\, (1+z_{A1835})^{3}}{3[\log(1+c)-c/(1+c)]}\,\frac{R^3}{r(r+R)^2} . \nn\\
\EEQ
From any mass density $\rho$, the tested quantity is
 \BEQ\label{SigmaBar=}
\overline\Sigma(r)
&=&\frac{4}{r^2}\int_0^r\d s\,s^2\rho(s)+\int_r^\infty\d s\,\frac{4s\rho(s)}{s+\sqrt{s^2-r^2}}.
\EEQ

As best fit to $\chi^2(\Sigmab)$ we find for NFW with $n_8=5$
\BEQ
A= 0.4330 \pm 0.0088  {m_N}/{\cm^3}, \quad R= 159.0 \pm 1.9\, \kpc .
\EEQ
Using  $z_{A1835}=0.253$ this corresponds to concentration $c=9.55\pm0.08$. 
With $\nu=17-2$, $\chi^2/\nu=5.5$ and $q=2.0\,10^{-11}$,  the case is formally ruled out at 6.7 $\sigma$.

The generalization ``gNFW'' involves a power $n\neq 1$  \cite{jing2002triaxial},
\BEQ
\rho_{\rm gNFW}=\frac{AR^3}{r^n (r+R)^{3-n}} . 
\EEQ
The  best gNFW fit again occurs for $n_8=5$,

\BEQ
A=0.2976 \pm 0.067  {m}_N/{\cm^3},\, \qquad  R=180 \pm 19\, \kpc ,
\EEQ
and $n=1.135 \pm0.036$, so that $c=8.18\pm0.72$.
This fit has $\chi^2/\nu=5.8$, $q=1.6\,10^{-11}$ and is formally ruled out at  6.8 $\sigma$.
The unexpected value $n>1$ is caused by the small errors of the data around 100 kpc, see fig. 1.
They arise since the lensing arclets produce mass maps with nearly the same $M_{2d}$ there.
For small $r$, on the other hand, there are fewer arclets and larger errors, while for large $r$ the relative errors increase as usual.

\section{Thermal particles} 

\subsection{Generalities}
We turn to thermal bosons for $g$ spieces of mass $m$ and chemical potential $m\mu$
at temperature $m\sigma^2$.
Setting $p=mv$, the Bose-Einstein mass density reads
\BEQ \label{ALPdens}
\rho_\DM(r)\!=\!\!\int \! \!\frac{{\rm d}^3v}{(2\pi\hbar)^3}\frac{gm^4}{\exp\{\,[\,\frac{1}{2}v^2+\varphi(r)-\mu]/\sigma^2 \}-s},
\hspace{2mm}
\EEQ
with $s=1$. For $s=0$ this describes isothermal classical particles and for $s=-1$ thermal fermions.

The data for the X-ray gas in A1835 fit well to   \cite{nieuwenhuizen2017modified}
\BEQ \label{rhogas=}
\rho_g(r)=\frac{\sigma_g^2(r^2+R_{g0}^2)}{2\pi G(r^2+R_{g1}^2)(r^2 + R_{g2}^2)}, 
\EEQ
with $\sigma_g=496.6\pm 6.4 $ km/s;  $\{R_{g0},R_{g1},R_{g2}\}= \{91 \pm 13, \, 31.8\pm  2.9, \, 169 \pm 15  \}$ kpc.
We model the galaxy mass density as  \cite{limousin2007combining}
\BEQ\label{rho-Gal}
\rho_\Gal(r)=
\frac{\rho_G^0}{(1+r^2/R_c^2)(1+r^2/R_t^2) } .
\EEQ
Solving the Poisson equation (\ref{Poissoneqn}) we may now determine $\Sigmab$ from (\ref{SigmaBar=}), 
which can also be expressed as \cite{nieuwenhuizen2009non}
 \BEQ\label{SigmaBarphi=}
\overline\Sigma(r)
=\frac{1}{\pi G}\int_0^\infty{\rm d} s\, \varphi'(r\cosh s) .
 \EEQ
With $\varphi'>0$ and varying less than $\rho$, this relation is numerically better behaved.

\subsection{Isothermal classical particles or objects}
Minimizing $\chi^2(\Sigmab)$ with respect to the free parameters in  (\ref{ALPdens}) and (\ref{rho-Gal}) we have $\nu=12$.  
Treating  $m^4$ and $\sigma^2$ as independent Gaussian variables, we obtain for the case $n_{10}=5$ 
\BEQ \label{parsMB}
&& m=  4.07^{+93}_{-4.07}\,g^{-1/4}\,e^{-\mu/4\sigma^2}{\eV}/{c^2},\, \, \,
\sigma=1464^{+2370}_{-1464}  \, \km/\s,  \\
&& \rho_c^0=42 \pm 2512 \frac{m_N}{\cm^3} , \,\,
\{R_c,R_t\}=\{1.5 \pm 45.4, \, 122 \pm 1151\} \,\kpc . \nn
\EEQ
The large error estimates and its $\chi^2/\nu=6.05$ express that the fit is bad.
It corresponds to $q=1.0\,10^{-10}$ and being formally ruled out at 6.5$\sigma$.

\subsection{Thermal bosons} 

\myskip{
We turn to thermal bosons for $g$ spieces of mass $m$ and chemical potential $m\mu$
at temperature $m\sigma^2$.
Setting $p=mv$, the Bose-Einstein mass density reads
\BEQ \label{ALPdens}
\rho_\DM(r)\!=\!\!\int \! \!\frac{{\rm d}^3v}{(2\pi\hbar)^3}\frac{gm^4}{\exp\{\,[\,\frac{1}{2}v^2+\varphi(r)-\mu]/\sigma^2 \}-s},
\hspace{2mm}
\EEQ
with $s=1$. For $s=0$ it describes isothermal classical particles and for $s=-1$ thermal fermions.

The data for the X-ray gas in A1835 fit well to   \cite{nieuwenhuizen2017modified}
\BEQ \label{rhogas=}
\rho_g(r)=\frac{\sigma_g^2(r^2+R_{g0}^2)}{2\pi G(r^2+R_{g1}^2)(r^2 + R_{g2}^2)}, 
\EEQ
with $\sigma_g=496.6\pm 6.4 $ km/s;  $\{R_{g0},R_{g1},R_{g2}\}= \{91 \pm 13, \, 31.8\pm  2.9, \, 169 \pm 15  \}$ kpc.
We model the galaxy mass density as  \cite{limousin2007combining}
\BEQ\label{rho-Gal}
\rho_\Gal(r)=
\frac{\rho_G^0}{(1+r^2/R_c^2)(1+r^2/R_t^2) } 
+\frac{\rho_G^1}{(1+r^2/R_u^2)^{n_c}} ,\qquad n_c=3.
\EEQ
Solving the Poisson equation (\ref{Poissoneqn}) we can now determine \cite{nieuwenhuizen2009non}
 \BEQ\label{SigmaBar=}
\overline\Sigma(r)
&=&\frac{4}{r^2}\int_0^r\d s\,s^2\rho(s)+\int_r^\infty\d s\,\frac{4s\rho(s)}{s+\sqrt{s^2-r^2}} \nn\\
&=&\int_0^\infty{\rm d} s\, \frac{\varphi'(r\cosh s)}{\pi G}.
 \EEQ
Fermion fits to data sets of A1689 have been 
reported \cite{nieuwenhuizen2009non,nieuwenhuizen2013observations,nieuwenhuizen2016dirac}. 
{\it Thermal bosons}. }

Let us return to the BE case (\ref{ALPdens})
for axions, \ALP s and dark photons. It is instructive to minimize $\chi^2(\Sigmab)$ for $n_{10}=5$  at fixed $\mu$, so that  $\nu=12$. 
The worst case occurs at $\mu=0$, 
\BEQ \label{parsBE}
&&
\hspace{-0mm}
m= 7.6^{+1.6}_{-7.6} g^{-1/4}{\eV}/{c^2},\quad
\sigma=1210^{+2000}_{1210}\, {\km}/{\s},    \\
&& 
\hspace{-0mm}
 \rho_0=286 \pm  1520\,  \frac{\mNbar}{\cm^3} ,\,\,
\{R_c,R_t\}=\{ 1.2\pm 3.4,  120\pm 380\} \, \kpc.  \nn
\EEQ
Its $\chi^2/\nu=12.5$ and $q=5.8\, 10^{-26}$ mean formal ruling out at $10.6\, \sigma$.
For $\mu$ taking increasingly negative values, $\chi^2$ diminishes untill
for $\mu\ll -\sigma^2$  the BE distribution approaches a MB one, with its large $m$ from (\ref{parsMB}).
Hence minimizing $\chi^2$ as function of $\mu$ will drive the best boson fit 
towards the classical isothermal limit, where it is still formally ruled out at  $6.5\sigma$.

\subsection{ Thermal fermions}

 After all these negative findings, we test eq. (\ref{ALPdens}) for fermions.
 Successful fermion fits to data sets of the cluster A1689 have been 
reported \cite{nieuwenhuizen2009non,nieuwenhuizen2013observations,nieuwenhuizen2016dirac}. 
For A1835 this case again yields a good fit. For $n_{10}=5$ the value  $\chi^2(\Sigmab)/\nu=1.82$ with $\nu=11$ and $q=0.046$
is  perfectly acceptable and proves the adequacy of our approach.
The  parameters are
\BEQ
\sigma&=&1164\pm {39} \, \km/\s , \qquad 
\mu= 5.8 \pm 1.3  \,10^6 \, \km^2/\s^2 ,  \nn \\
\rho_G^0&=& 18 \pm  41\, m_N /\cm^3 , \\ 
R_c&=&7.2 \pm 9.2 \, \kpc ,\qquad
R_t=123 \pm  160\, \kpc.  \nn
\EEQ
For the DM density the parameters are reasonably constrained, but for the galaxies not. The mass takes the value
\BEQ \label{mnu=}
m&=&1.61^{ + 0.19}_{ - 0.30} \,\Big( \frac{12}{g}\Big)^{1/4} \,  \eV/c^2.
\EEQ

The fit is presented in fig. 1. The residues have a systematic trend, again induced by the small errors around 100 kpc
and minimized by choosing bin 10 as the one with 5 points.

In our approach the dark matter is fitted together with the galaxies. One may wonder whether this induces a bias towards fermions.
However, dropping the galaxies mass density and only fitting the last 6 bins again brings fermions as best fit, be it with mass of 
$2.14\,(12/g)^{1/4} $ eV.

\subsection{Interpretation in terms of neutrinos}

The fermionic case likely refers to neutrinos and anti-neutrinos. 
Indeed, they act as  $g/2$ relativistic degrees of freedom during the BBN, 
which poses new issues, so it is economic that some of them are known particles.

Active neutrinos are in principle Majorana particles, but with eV mass, neutrinoless double $\beta$-decay should 
have been discovered. Indeed, GERDA  gives as most recent result 
$m_{\beta\beta}^{0\nu}<0.15$ -- $0.33$ eV \cite{gerda2017background},
where, in the usual notation \cite{lesgourgues2013neutrino},
\BEQ
m_{\beta\beta}^{0\nu}\equiv |c^2_{12} c^2_{13}m_1 +e^{i\eta_1}s^2_{12} c^2_{13}m_2 +e^{i\eta_2} s^2_{13} m_3|. 
\EEQ
For equal $m_{1,2,3}=m_\nu$ and $\eta_{1,2}=\pi$, the known mixing angles \cite{lesgourgues2013neutrino}
yield the value $0.37 m_\nu$, so that in general $m_\nu\le 2.8\,m_{\beta\beta}^{0\nu}$. 
Violating this bound for any $g\lesssim 110$, our neutrinos must be of (nearly) Dirac type  \cite{nieuwenhuizen2016dirac}.
Up to the small effects of neutrino oscillations, the active neutrinos have (nearly) equal mass, also 3 sterile 
partners with (nearly) this mass and a  (nearly) zero sterile Majorana mass matrix
\cite{lesgourgues2013neutrino}. 
With the antineutrinos there are $g=12$ fermion species or 3 + 3 fermion families.

The number density is $56$ $\cm^{-3}$ for each species \cite{lesgourgues2013neutrino}, so if
the cold dark matter fraction $\Omega_c$ actually stems from neutrinos, the WMAP value  \cite{hinshaw2013nine}
corresponds to $m=1.80\pm0.08$ eV and the Planck value \cite{ade2016planck} to  $m=1.88 \pm 0.03$ eV.
DES Y1  \cite{troxel2017dark} implies $m= 1.68_{-0.15}^{+ 0.25} h_{70}^2$ eV.
Within 1.5$\sigma$ these cases are covered by (\ref{mnu=}) and support our findings for A1689  \cite{nieuwenhuizen2009non,nieuwenhuizen2013observations,nieuwenhuizen2016dirac}.

 \begin{figure}
\label{SiBA1835}
\includegraphics[scale=0.9]{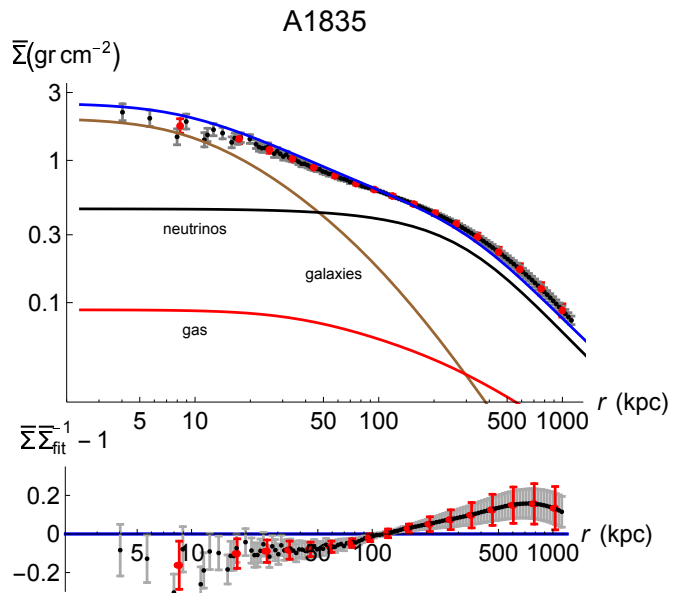}
\caption{Data for $\Sigmab$ in A1835 (black points, gray error bars) and binned data (red).
Upper line: best fit for thermal fermion model (blue). 
Lower lines: contributions from neutrinos, galaxies and X-ray gas, respectively. Lower pane: fit residuals. }
\end{figure}

\section{Summary}

After recalling that modifications of Newton's law do not solve the dark matter problem 
in galaxy clusters  \cite{nieuwenhuizen2017zwicky,nieuwenhuizen2017modified}, 
we consider the performance of the most studied DM candidates in clusters.
An important question is whether DM is self-interacting (SiDM). If this is indeed the case, its
elastic cross section $\sigma/m\sim 2\, \cm^2/\gr$ puts strong constraints: \MACHO s and primordial black holes 
are ruled out, together with light axion-like particles that have to build \MACHO s. 
It would also put constraints on other particle models,
for instance, axions and sterile neutrinos should, at best, scatter very weakly. 
Hence the establishment or ruling out of SiDM in cluster collision is of major interest. 

In contrast to CMB and BAO analyses, relaxed clusters provide a simple cosmological test, 
because their history has just led to a certain relaxed shape for the DM and can be disregarded.
To compare to our previous works on A1689, we consider here the cluster A1835, for which strong lensing and X-ray data 
were presented  \cite{nieuwenhuizen2017modified}.
We introduce a new, parameter-free method to regularize the small eigenvalues of the covariance matrix:
binning and accounting for the intra-bin variations. 
We present results for one particular way of binning and fitting; other ones produced the same trend.
Within this approach we analyze several options for dark matter. 
NFW models and classical isothermal models do not fare well 
for the small errors  in the data 
and seem eliminated at more than $6\sigma$;
hence even if DM turns out not to be self-interacting, \MACHO s and \PBH s seem to be ruled out.
Thermal bosonic models perform even less well unless they are in their classical isothermal limit;
this severely questions whether thermal axions or \ALP s can constitute the DM.

Thermal fermionic DM, however, does offer a good match.
They have to represent (nearly) Dirac neutrinos with a mass of 1.5 -- 1.9 eV;
also the 3 right handed sterile partners  have (nearly) this mass and 
a (nearly) vanishing Majorana mass matrix.
The exclusion of more than one sterile neutrino in oscillation experiments  \cite{giunti2016light} would not concern them.

If neutrinos indeed have a such a large mass, nonlinearities will be needed in the plasma phase
to circumvent the free-streaming road block of linear structure formation. 
But the notorious the $^7$Li problem in the BBN may as well require nonlinearities.
The latter could be restricted to the cluster scale and down to the galaxy scale or lower, and have not much impact on the CMB.
Neutrinos with eV mss have no impact inside galaxies, but the solution could lie in MOND \cite{milgrom1983modification} 
or gravitational hydrodynamics \cite{nieuwenhuizen2009gravitational}.

An effect similar to dark matter self-interaction in cluster-cluster collision may be caused by the Pauli principle acting
in the collision of such quantum degenerate ``neutrino stars''.

\section{Outlook}  

The question raised by previous studies of A1689 and now confirmed for A1835 becomes pressing: What is the reason for 
singling out degenerate fermions as best fit for cluster lensing? 
While it is desirable to study more clusters, preferably relaxed spherical ones, one may already wonder:
Is there a conspiracy,  or is simply the neutrino, after all, just the dark matter particle, 
and \LCDM \, only an effective theory? And is the neutrino a Dirac fermion just having its right handed partner?
The answer will come from the  test of the electron antineutrino mass in the KATRIN experiment  \cite{ottenweinheimer2008};
for the  prediction of  1.5 -- 1.9 eV  two months of data taking in 2018   \cite{cho2017weighing} should suffice.
 If such a detection is indeed made, the neutrino sector of the standard model is basically determined and the cluster dark matter riddle solved.

{\it Acknowledgments} We thank A. Morandi, M. Limousin and E.F.G. van Heusden for discussion.

\bibliographystyle{prop2015} 
 \bibliography{A1689-dragging-bib}

\end{document}